\documentclass[aip,pop,reprint]{revtex4-1}
\usepackage{hyperref}
\usepackage{float,amsmath,amssymb,amsfonts,graphicx,natbib}
\usepackage{dcolumn,array}
\usepackage[page]{appendix}
\newcommand{\be}{\begin{equation}}
\newcommand{\ee}{\end{equation}}

\DeclareMathOperator{\diag}{diag}
\usepackage{setspace}

\begin{document}
\title{Subcritical turbulence spreading and avalanche birth}
\author{R.A. Heinonen}
\email{robin@physics.ucsd.edu}
\affiliation{University of California, San Diego, California 92093, USA}
\author{P.H. Diamond}
\affiliation{University of California, San Diego, California 92093, USA}
\date{\today}

\begin{abstract}
In confined plasmas, a localized fluctuation in a marginal or weakly damped region will propagate and generate an avalanche if it exceeds a threshold. In this letter, a new model for turbulence spreading based on subcritical instability in the turbulence intensity is introduced. We derive a quantitative threshold for spreading from a seed in a stable region, based on a competition between diffusion and nonlinear growth of the turbulence intensity. The model resolves issues with the established Fisher equation model for turbulence spreading, which is supercritical and cannot support the stationary coexistence of multiple turbulence levels.  Implications for turbulence spreading are discussed, including the dynamics of ballistic penetration of turbulence into the stable zone. Tests of the theory are suggested.
\end{abstract}

\maketitle

The paradigm of directed percolation \cite{hinrichsen} is ubiquitous in non-equilibrium statistical dynamics. Directed percolation is realized by the contamination and ultimate excitation of, say, a localized nonlinear oscillator by interaction with its neighbors \cite{pomeau86}. For the contaminated region to expand, the interaction must be strong enough to excite neighbors against their damping. This process of expansion of excited regions by directed percolation is relevant to many problems in the spatiotemporal dynamics of turbulence \cite{pomeau15}. These include, but are not limited to, the expansion (and ultimate overlap) of turbulent slugs at the onset of turbulence in high Reynolds number pipe flow turbulence \cite{barkley,sipos}, the spreading of a turbulent patch \cite{barenblattbook}, and the Loitsyansky problem \cite{davidson} for large-scale evolution of 3D turbulence. The theme of \emph{spreading-by-contamination} \cite{TS18} appears in the magnetic confinement physics phenomena of \emph{turbulence spreading} \cite{garbet94,hahm04,linhahm,naulin} and \emph{avalanching} \cite{DH95,politzer,sarazin}. These closely-related phenomena are of pragmatic interest in that they delocalize the flux-gradient relation which governs turbulent transport. This letter applies the physics of spreading-by-contamination to propose a mechanism for how turbulence penetrates stable regions. 

Indeed, experiments and simulations of magnetic fusion (MF) plasma strongly suggest that the transport can be \emph{nonlocal}: fluxes cannot be determined by local parameters and their gradients, but are generally given by some integrated spatiotemporal relation that depends on global profile structure (not only local gradients) \cite{ida,inagaki}. Such phenomena cannot be explained by diffusive or Fickian processes and instead depend on fast propagation dynamics of turbulence in the plasma. Turbulence spreading and avalanching, both of which involve radial self-propagation of turbulence on mesoscales, are two players which influence nonlocal transport. Turbulence spreading, arguably the more general phenomenon, is the result of nonlinear coupling between ballooning modes, which results in spatial scattering and nonlinear diffusion of the turbulence energy. Standard spreading problems are the spatiotemporal evolution of a single, initially localized spot of turbulence, and the penetration of turbulence into a linearly stable region. An avalanche refers to a bursty transport event which propagates from an initial ``seed'' (some large, localized fluctuation) via the sequential, domino-like overturning of cells, coupled through local gradients. Avalanching exhibits many features of self-organized criticality (SOC), such as $1/f$ noise \cite{sarazin}.

Spreading and avalanching share several common features. Both appear on a broad range of mesoscales $\Delta_c \ll \ell \ll a,$ i.e. are excitations longer than the correlation length of the microturbulence but much shorter than the system size. Both are fast relative to transport timescales but slower than the microturbulence correlation time. Both originate from three-mode coupling in wavenumber space. Also, both result in nonlinear, ballistically propagating fronts/pulses of turbulence, for which $\Delta x \sim t$. Therefore, it is desirable to describe spreading and avalanching by a unified reduced model. Such a model should not only capture the relatively well-understood superdiffusive turbulent front propagation, but also should help us understand under what conditions an avalanche will occur, i.e.\ it should predict the size and scale of the minimal seed. 

The conventional approach \cite{gurcan05,hahm04,naulin} to modeling turbulence spreading is to use a simple, 1D equation for the turbulence intensity field, in the spirit of a $K$-$\epsilon$ model. Specifically, a reaction-diffusion equation of Fisher(-KPP) type is employed:
\be
\label{eq:fisher}
\partial_t I = \gamma_0 I - \gamma_{NL} I^2 + \partial_x \left(D_0 I \partial_x I\right).
\ee
The first term on the RHS represents local linear turbulence drive, with $\gamma_0$ the growth or damping rate. The second term results in nonlinear saturation of the intensity at the mixing length level $I=\gamma_{NL}/\gamma_0,$ when $\gamma_0>0$. The nonlinear damping $\gamma_{NL}$ includes effects due to zonal flow shearing and mode-mode coupling. The third and final term is the nonlinear diffusion of the turbulence energy due to mode-mode coupling, which results in spatial scattering. This arises by closing the $E\times B$ convective nonlinearity and employing an envelope approximation, yielding \cite{kim}
\begin{align}
\sum_{\mathbf{k}'} (\mathbf{k} \cdot \mathbf{k}' \times \hat{z})^2 | \tilde{\phi}_{\mathbf{k}'}|^2 R(\mathbf{k},\mathbf{k}') I_{\mathbf{k}} &\to - \frac{\partial}{\partial x} D_x(I) \frac{\partial}{\partial x} I_{\mathbf{k}}\\  \nonumber
& + \mathbf{k}\mathbf{k} : \mathbf{D} I_{\mathbf{k}},\label{eq:diffusion}
\end{align}
where $R$ is a resonance function which determines the correlation time, $D_x = \sum_{\mathbf{k}'}  k_y'^2| \tilde \phi_{\mathbf{k'}}|^2 R(\mathbf{k},\mathbf{k}'),$ and $\mathbf{D} = \sum_{\mathbf{k}'} \diag(k_x'^2,k_y'^2) | \tilde{\phi}_{\mathbf{k}'}|^2 R(\mathbf{k},\mathbf{k}'). $ (Here, $\diag(a,b)$ refers to the $2\times2$ diagonal matrix with elements $a$ and $b$.) The first term is the (radial) nonlinear diffusion and the second represents local nonlinear transfer. We note that other forms (say, $\propto I^\alpha$) of the diffusion are possible, but the choice $D(I) = D_0 I$ is standard and simple. In principle, the coefficients $\gamma_0$, $\gamma_{NL}$, $D_0$ are spatially varying, but for simplicity we will generally take them to be uniform. 

In the Fisher model, turbulence spreading is realized by propagating fronts which connect the turbulent and laminar fixed points. However, such front solutions exist only in linearly unstable regions, where the presence of any noise should have already excited the system to turbulence. Thus, the fronts are manifestly unphysical! Moreover, the Fisher model predicts very weak, evanescent penetration of turbulence into linearly stable regions, which is dubious in light of clear experimental observations of fluctuations occurring in such zones \cite{nazikian}. 

In this letter, then, we introduce a new, phenomenological model based on \emph{bistability} of the turbulence intensity. This model exhibits several desirable features in common with the Fisher model, while simultaneously providing a basic framework with which to understand bursty, avalanche-like events. It involves a \emph{subcritical} bifurcation which allows robust coexistence of turbulent and laminar regions (even in the presence of noise)---something which is impossible in the supercritical Fisher model. We proceed as follows: first, the conventional wisdom on turbulence spreading is discussed. Second, we introduce the new bistable model and justify it with physical arguments. Third, the new features of this model, and especially its implications for turbulence spreading, are discussed. Finally, we derive the threshold size for an initial, localized seed of turbulence to trigger front propagation in a stable region. The key physics governing this threshold is a competition between turbulence intensity scattering and local nonlinear growth. Such a transport event originating from a seed is similar to an avalanche, but propagating by scattering rather than gradient coupling, which is neglected in this model.

Our model rests on the assumption of subcritically unstable/bistable turbulence, similar to the well-known cases of Poiseuille and plane Couette flow \cite{orszag}. There is considerable evidence suggesting that the turbulence in MF plasma may indeed be subcritical, as well. Inagaki \emph{et al.}\ \cite{inagaki} have demonstrated \emph{global hysteresis} between the turbulence intensity and the local temperature gradient in the Large Helical Device. This hysteresis, which occurs in the absence of a discernible transport barrier, implies a memory effect in the turbulence and suggests the possibility of a bistable, S-curve relation between the intensity and the local gradient. From theory and simulation, the 3D Hasegawa-Wakatani system is known to be subcritical in the presence of magnetic shear, due to a nonlinear streaming instability \cite{biskamp,drake,friedman15}. Self-organized, self-sustaining turbulence has also been observed in simulations of a more complex sheared system based on the Braginskii equations \cite{scott90,scott92}. Guo and Diamond \cite{GD17} argued that the turbulence exhibits bistability in the vicinity of temperature corrugations in the $E\times B$ staircase, a self-organized structure of quasiperiodic shear flow layers \cite{dif2010}. These corrugations arise due to inhomogeneous turbulent mixing and drive ITG turbulence locally, which can result in further roughening of the temperature profile and thus nonlinear turbulence drive. A sufficiently strong ambient perpendicular shear flow can also cause the turbulence to be subcritical \cite{CSD92,pringle}---this effect has been observed in gyrokinetic simulations \cite{barnes,vanwyk}. Finally, we note that phase-space structures such as holes and granulations, which can interact with drift waves \cite{holes}, are known to drive nonlinear instabilities \cite{lesur}. 

In light of the above considerations, we propose the following minimal model for the spreading of subcritical turbulence:
\be
\label{eq:rd}
\partial_t I = f(I) + \partial_x \left(D(I) \partial_x I\right) + \xi(x,t)
\ee
with cubic reaction function
\be
\label{eq:cubic}
f(I) = \gamma_1 I + \gamma_2 I^2 - \gamma_3 I^3
\ee
and $\gamma_2,\gamma_3>0.$ $\xi(x,t)$ is noise, which we ignore for now. We again take $D(I) = D_0 I$. This can be viewed as a simple extension of the earlier paradigm---the term $\propto I$ is the local linear drive, the term $\propto I^3$ is local nonlinear saturation, and we again have nonlinear diffusion. The new physics is contained in the term $\propto I^2$, which represents nonlinear instability. In a certain regime this gives rise to bistability between a laminar state (or more generally, a background turbulence level) and a saturated turbulence level. A rigorous derivation of the coefficients from a full nonlinear model for the plasma evolution would require facing the difficult task of resolving the large-amplitude nonlinear physics, well beyond the breakdown of quasilinear theory. Instead, we note that this model captures the \emph{qualitative} dynamics of any model for spreading of bistable turbulence. One might anticipate drift-wave scalings $D_0 \sim \chi_{GB}$, $\gamma_1 \sim \epsilon \omega_*,$ and (invoking critical balance) $\gamma_2\sim \gamma_3 \sim \omega_*$. Here $\epsilon$ represents the deviation from linear marginality. Our model is similar to some models for the pipe flow transition to turbulence \cite{barkley,pomeau15}, as well as Gil and Sornette's model for avalanching in sandpiles \cite{gilsornette}.

Eq.~(\ref{eq:rd})--(\ref{eq:cubic}) may be written in variational form
\be
\label{eq:variational}
D(I) \partial_t I = - \frac{\delta {\cal F}}{\delta I}
\ee
with Lyapunov functional
\be
\label{eq:functional}
{\cal F}[I] \equiv \int dx \, \left[ \frac{1}{2} (D(I) \partial_x I)^2 - \int_0^I dI' \, D(I') f(I')\right],
\ee
where $\frac{d{\cal F}}{dt} \le 0$ at all times. The extrema of the ``potential part'' ${\cal V}(I) \equiv - \int dx \, \int_0^I dI' \, D(I') f(I')$ then determine the uniform fixed points of the model and their stability. Of particular interest is the weak damping regime with $-{\gamma_2}^2/4\gamma_3<\gamma_1<0$. It is in this regime that the model is bistable, and the dynamics are qualitatively different from those of the Fisher model (see Table~\ref{table:regimes} for details of the various parameter regimes). In particular we may define the fixed points $I=I_\pm = (\gamma_2 \pm \sqrt{{\gamma_2}^2+4\gamma_1 \gamma_3})/2\gamma_3,$ transform $I \to I_+ I,$ and write $|\gamma_3| I{_+}^2 \to \gamma, \;
 \frac{I_-}{I_+} \to \alpha, \; {I_+} D_0 \to D$ to yield the Zel'dovich-Frank-Kamenetsky equation \cite{ZFK}
\be
\label{eq:nagumo}
\partial_t I = \tilde{f}(I) + \partial_x(\tilde{D}(I) \partial_x I),
\ee
with $\tilde{f}(I) = \gamma I (1-I)(I-\alpha)$ and $\tilde{D}(I) = D I$ (we will drop the tildes henceforth).  The remaining dimensional parameters $\gamma$ and $D$ can also be scaled out of the problem via the transformation $x\to \sqrt{D/\gamma} x$ and $t \to t/\gamma$, which provides the natural length and time scales of the model. 

\begin{table*}
\caption{\label{table:regimes} Summary of features of the various parameter regimes in the cubic model. Unphysical roots ($I<0$) are ignored.}
\begin{ruledtabular}
\begin{tabular}{lllll}

Regime & Stable roots & Unstable roots & Fronts? & Comments \\
\hline
$\gamma_1 > 0$ &  $I_+$ & 0 & forward-propagating & similar to Fisher with $\gamma_0>0$\\
\hline
$\gamma_1 <0$, $|\gamma_1|\gamma_3/\gamma_2^2 < \frac{15}{64}$ & $0, I_+$ & $I_-$ & forward-propagating & $\alpha<\alpha^*$; turbulent root abs. stable \\
\hline
$\gamma_1 <0$, $\frac{15}{64} <|\gamma_1|\gamma_3/\gamma_2^2 < \frac14 $ & $0, I_+$ & $I_-$ & receding & $\alpha>\alpha^*$; turbulent root metastable \\
\hline
$\gamma_1 <0,$ $|\gamma_1|\gamma_3/\gamma_2^2 > \frac14 $ & 0 & none & none & similar to Fisher with $\gamma_0<0$
\end{tabular}
\end{ruledtabular}
\end{table*}

Threshold behavior is now apparent: above $I=\alpha$ (and below $I=1$), the nonlinear instability overcomes the linear and cubic damping terms, and the turbulence grows locally. This corresponds to a potential barrier in ${\cal V} (I)$ at $I=\alpha.$ This threshold behavior can lead to hysteresis, similar to that observed in Inagaki \emph{et al.}, if the turbulence level is raised/lowered globally. As we will see, the nonlinear threshold is also symptomatic of avalanching.

The bistable regime supports traveling turbulence front solutions with characteristic lengthscale $\sqrt{D/\gamma}$ which are qualitatively similar to those of the Fisher model, yet propagate in a linearly stable or marginal region \cite{SGM97}. The propagation speed $c$ obeys the Maxwell construction 
\be
\label{eq:maxwell}
c \int_{-\infty}^{\infty} D(I(z)) I'(z)^2 \, dz = \int_0^1 D(I) f(I) \, dI
\ee
where $z=x-ct$ and we have set $I(z=-\infty)=1$ and $I(z=\infty)=0.$ In particular, the sign of $c$ is given by the sign of the integral on the RHS, and thus turbulence spreads when $\alpha<\alpha^* = 3/5,$ recedes when $\alpha > \alpha^*$, and is stationary for $\alpha=\alpha^*.$ Note that $I=0$ is a metastable minimum of $V(I)$, and $I=1$ is absolutely stable, when $\alpha<\alpha^*$; the reverse is true when $\alpha>\alpha^*$.  

A prediction of this model which differs strongly from the corresponding Fisher model prediction concerns the propagation of turbulence from an unstable zone into a stable zone. This problem is, for example, relevant to the turbulent contamination of transport barriers \cite{ghendrih}. This can be modeled by fixing a linear growth rate $\gamma_g>0$ for $x<0$ and a linear damping rate $\gamma_d$ for $x>0$ and allowing a turbulence wavefront to develop in the lefthand region. In the Fisher case, the turbulence barely penetrates into the stable zone, forming a stationary, exponentially decaying intensity profile with characteristic depth 
$\lambda \sim \sqrt{\frac{D_0}{\gamma_{NL}}} + O\left(\log\frac{\gamma_g}{\gamma_d}\right),$ assuming $\gamma_g\gtrsim\gamma_d$ \cite{gurcan05}. However, the subcritical model allows for the possibility of \emph{ballistic} turbulence propagation into the stable zone. In particular, one easily sees that if the damping is so weak that $\alpha<\alpha^*$, a new propagating wave is formed in the stable region, albeit with reduced amplitude and speed. This results in significantly stronger delocalization of the flux-gradient relation than is possible in a supercritical model.

We now turn to the problem of spreading from a seed. In the Fisher model, all fluctuations in a stable zone damp like $\exp(-|\gamma_0| t)$. However, for subcritical turbulence, a \emph{sufficiently large} spot or puff of turbulence, initially localized, will grow and expand. This is akin to an avalanche. (The spreading of a turbulent spot is also a classical problem in fluid mechanics.) We will focus on the case $\alpha<\alpha^*,$ but note that if $\alpha>\alpha^*$ there is the interesting possibility of an ``inverse avalanche'' triggered by a local \emph{depression} in the turbulence intensity, i.e. receding fronts with $c<0$ in Eq.~(\ref{eq:maxwell}).

\begin{figure}
\includegraphics[width=\columnwidth]{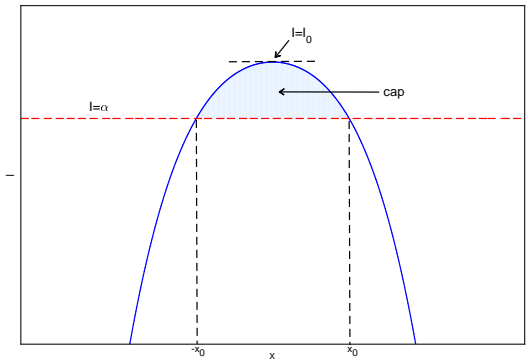}
\caption{The ``cap'' of a puff of turbulence. The competition between nonlinear turbulence growth in the cap and diffusive flux out of the cap at $x=\pm x_0$ generates a threshold lengthscale for growth of the initial condition.}
\label{fig:cap}
\end{figure}

A puff of turbulence must exceed the nonlinear instability threshold somewhere (i.e.\ $I>\alpha$) in order to propagate. The puff must also exceed a threshold \emph{spatial scale}. To see this, consider the ``cap'' of the initial profile, i.e.\ the part exceeding $I=\alpha,$ as shown in Figure \ref{fig:cap}. The threshold is determined by a competition between the outgoing diffusive flux from the cap, and the total nonlinear growth of the turbulent intensity within the cap. The turbulent mass scattered out of the cap will enter a region where the effective growth is negative and the turbulence will dissipate. This competition is also suggested by the form of the Lyapunov functional, Eq.~(\ref{eq:functional}). $\sqrt{D/\gamma}$ then sets a lengthscale for the minimum seed size. Interestingly, diffusion both drives the ``avalanche'' by providing the spatial coupling mechanism and limits it by increasing the threshold size. 

We now determine the dependence of the threshold size on amplitude. Assume for simplicity that the initial profile $I(x)$ of the puff is smooth and even, and that it has a single associated lengthscale $L$ and a single maximum $I_0$ at $x=0.$ Define $L$ so that $I''(0) = - I_0/L^2$ (we assume this derivative is nonzero). In the cap, we have
\be I(x) = I_0 - \frac{ I_0}{2L^2} x^2 +O(x^4)\ee
and $I(x)$ crosses $I=\alpha$ at $\pm x_0,$ where $x_0 \simeq \sqrt{2 \left(1-\frac{\alpha}{I_0}\right)} L.$ We integrate Eq.~(\ref{eq:nagumo}) from -$x_0$ to $x_0.$ Expanding
\be
 \int_{-x_0}^{x_0} dx \, f(I) \simeq  2x_0 f(I_0) - \frac{ I_0}{3L^2} {x_0}^3 f'(I_0), \ee
one finds that the turbulent mass in the cap, ${\cal I} \equiv \int_{-x_0}^{x_0} dx \, I,$ evolves as 
\be
\partial_t {\cal I} \simeq 2 f(I_0) x_0 - \frac{ I_0}{3L^2} {x_0}^3 f'(I_0) -  \frac{2 D (\alpha) x_0}{L^2},
\ee
The mass in the cap thus grows initially if
\be
\label{eq:thresh}
L \gtrsim L_{\rm min} = \sqrt{\frac{3 D\alpha I_0}{\gamma (I_0-\alpha) ((1-2\alpha)I_0 + \alpha)}}.
\ee

\begin{figure}
\includegraphics[width=\columnwidth]{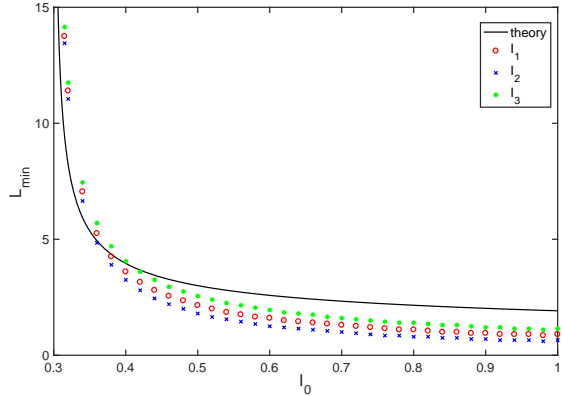}
\caption{Numerically obtained lengthscale threshold $L_{\rm min}(I_0)$ at $\alpha=0.3$ for three different functional forms for the initial data: $I_1=I_0 \exp(-x^2/L^2)$ (red), $I_2= I_0 /(1+x^2/L^2)$ (blue), and $I_3 = I_0 (1-\frac{x^2}{L^2})$ supported on $|x|<L$ (green). The numerical results are compared with Eq.~(\ref{eq:thresh}) (solid line). $L_{\rm min}$ is expressed in units of $\sqrt{D/\gamma}.$}
\label{fig:thresh}
\end{figure}

In particular, we find power law behavior $L_{\rm min} \sim (I_0-\alpha)^{-1/2}$ for $I_0\gtrsim \alpha$. The above estimate agrees well with numerical simulation of the PDE---see Fig.~\ref{fig:thresh}. The exponent $-1/2$ holds for any initial data close to threshold, does not depend on the form of the diffusion, and is robust to transformations in the reaction function of the form $I\to I^\beta$ ($\beta>0$). 

How might this threshold be exceeded in a real system? In the weak linear damping limit $|\gamma_1| \ll {\gamma_2}^2/\gamma_3,$  we have $I_-\sim \frac{|\gamma_1|}{\gamma_2}$ and $L_{\rm min} \sim \left(\frac{D_0}{\gamma_2}\right)^{1/2} \sim \left(\frac{\chi_{GB}}{\omega_*}\right)^{1/2} \sim \Delta_c,$ where $\Delta_c$ is the correlation length of the turbulence. $\frac{|\gamma_1|}{\gamma_2}$ is small in this limit, and $\Delta_c$ is essentially the smallest lengthscale of interest, so these are relatively weak constraints. This suggests the possibility of avalanche excitation by noise near the linear stability threshold. One possibility for the noise is $\xi(x,t) = (I(x,t)+I_0) \eta(x,t)$, where $\eta$ is Gaussian white noise. Here $I_0$ represents a small background, say due to sub-ion scale turbulence. The multiplicative noise $\propto I(x,t)$ is a simple, reasonable choice that (correctly) vanishes in the absence of fluctuations. Simulations of the stochastic PDE  (Eq.~(\ref{eq:rd})) initialized at $I=0$ indicate that small puffs of turbulence will spontaneously form and grow as a result of the multiplicative noise. As compared to linear diffusion $\propto \partial_{xx} I,$ the nonlinear diffusion is ineffective at smoothing out the puffs because the term $\propto I \partial_{xx} I$ vanishes at small amplitude. Close to marginality, puffs can exceed the threshold and form a propagating front. This is consistent with the bursty, intermittent character of avalanching. Note that the turbulence transition in pipe flow is similarly intermittent \cite{pomeau80}. 

In conclusion, this letter derives a threshold for turbulence spreading into a linearly stable region. The subcritical model for turbulence spreading considered above is a substantial improvement over the Fisher model. The new model is motivated by numerous theoretical arguments and observations in experiment and simulation which support subcritical turbulence, and it resolves the issue that supercritical models cannot reasonably support coexistence of multiple turbulence levels. It provides a simple framework for understanding how localized puffs of turbulence grow beyond the nonlinear stability threshold, thus triggering intermittent avalanching and spreading into stable regions.

The model is directly testable in at least two ways. First, it predicts a power-law estimate for the critical size for the seed of an avalanche, which can be compared to simulations. Moreover, it predicts the possibility of strong \emph{ballistic} invasion of turbulence into a stable zone with weak damping. Such behavior is forbidden in a supercritical model, and, if observed in simulation or experiment, would be suggestive of turbulence bistability. However, such behavior would be challenging to disambiguate from an avalanche proceeding by gradient propagation (which we artificially neglect), unless the gradient were to remain subcritical after the penetration of turbulence. Finally, the model may be tested by studies of the response to localized stimuli, as in recent basic experiments on avalanching \cite{compernolle}. 

Finally, we note that this picture is of course highly simplified; a complete model of spreading must include couplings to zonal flows and profiles, which is the subject of future work. 
\acknowledgements{We acknowledge Guilhem Dif-Pradalier, Zhibin Guo, and \"Ozg\"ur G\"urcan for useful discussions, as well as fruitful interactions at the 2018 Chengdu Theory Festival. This material is based upon work supported by the U.S. Department of Energy, Office of Science, Office of Fusion Energy Sciences under Award Number DE-FG02-04ER54738.}

%


\begin{thebibliography}{40}%
\makeatletter
\providecommand \@ifxundefined [1]{%
 \@ifx{#1\undefined}
}%
\providecommand \@ifnum [1]{%
 \ifnum #1\expandafter \@firstoftwo
 \else \expandafter \@secondoftwo
 \fi
}%
\providecommand \@ifx [1]{%
 \ifx #1\expandafter \@firstoftwo
 \else \expandafter \@secondoftwo
 \fi
}%
\providecommand \natexlab [1]{#1}%
\providecommand \enquote  [1]{``#1''}%
\providecommand \bibnamefont  [1]{#1}%
\providecommand \bibfnamefont [1]{#1}%
\providecommand \citenamefont [1]{#1}%
\providecommand \href@noop [0]{\@secondoftwo}%
\providecommand \href [0]{\begingroup \@sanitize@url \@href}%
\providecommand \@href[1]{\@@startlink{#1}\@@href}%
\providecommand \@@href[1]{\endgroup#1\@@endlink}%
\providecommand \@sanitize@url [0]{\catcode `\\12\catcode `\$12\catcode
  `\&12\catcode `\#12\catcode `\^12\catcode `\_12\catcode `\%12\relax}%
\providecommand \@@startlink[1]{}%
\providecommand \@@endlink[0]{}%
\providecommand \url  [0]{\begingroup\@sanitize@url \@url }%
\providecommand \@url [1]{\endgroup\@href {#1}{\urlprefix }}%
\providecommand \urlprefix  [0]{URL }%
\providecommand \Eprint [0]{\href }%
\providecommand \doibase [0]{http://dx.doi.org/}%
\providecommand \selectlanguage [0]{\@gobble}%
\providecommand \bibinfo  [0]{\@secondoftwo}%
\providecommand \bibfield  [0]{\@secondoftwo}%
\providecommand \translation [1]{[#1]}%
\providecommand \BibitemOpen [0]{}%
\providecommand \bibitemStop [0]{}%
\providecommand \bibitemNoStop [0]{.\EOS\space}%
\providecommand \EOS [0]{\spacefactor3000\relax}%
\providecommand \BibitemShut  [1]{\csname bibitem#1\endcsname}%
\let\auto@bib@innerbib\@empty
\bibitem [{\citenamefont {Hinrichsen}(2000)}]{hinrichsen}%
  \BibitemOpen
  \bibfield  {author} {\bibinfo {author} {\bibfnamefont {H.}~\bibnamefont
  {Hinrichsen}},\ }\href {\doibase 10.1080/00018730050198152} {\bibfield
  {journal} {\bibinfo  {journal} {Advances in Physics}\ }\textbf {\bibinfo
  {volume} {49}},\ \bibinfo {pages} {815} (\bibinfo {year} {2000})},\ \Eprint
  {http://arxiv.org/abs/https://doi.org/10.1080/00018730050198152}
  {https://doi.org/10.1080/00018730050198152} \BibitemShut {NoStop}%
\bibitem [{\citenamefont {Pomeau}(1986)}]{pomeau86}%
  \BibitemOpen
  \bibfield  {author} {\bibinfo {author} {\bibfnamefont {Y.}~\bibnamefont
  {Pomeau}},\ }\href {\doibase https://doi.org/10.1016/0167-2789(86)90104-1}
  {\bibfield  {journal} {\bibinfo  {journal} {Physica D: Nonlinear Phenomena}\
  }\textbf {\bibinfo {volume} {23}},\ \bibinfo {pages} {3 } (\bibinfo {year}
  {1986})}\BibitemShut {NoStop}%
\bibitem [{\citenamefont {{Pomeau}}(2015)}]{pomeau15}%
  \BibitemOpen
  \bibfield  {author} {\bibinfo {author} {\bibfnamefont {Y.}~\bibnamefont
  {{Pomeau}}},\ }\href {\doibase 10.1016/j.crme.2014.10.002} {\bibfield
  {journal} {\bibinfo  {journal} {Comptes Rendus Mecanique}\ }\textbf {\bibinfo
  {volume} {343}},\ \bibinfo {pages} {210} (\bibinfo {year}
  {2015})}\BibitemShut {NoStop}%
\bibitem [{\citenamefont {Barkley}\ \emph {et~al.}(2015)\citenamefont
  {Barkley}, \citenamefont {Song}, \citenamefont {Mukund}, \citenamefont
  {Lemoult}, \citenamefont {Avila},\ and\ \citenamefont {Hof}}]{barkley}%
  \BibitemOpen
  \bibfield  {author} {\bibinfo {author} {\bibfnamefont {D.}~\bibnamefont
  {Barkley}}, \bibinfo {author} {\bibfnamefont {B.}~\bibnamefont {Song}},
  \bibinfo {author} {\bibfnamefont {V.}~\bibnamefont {Mukund}}, \bibinfo
  {author} {\bibfnamefont {G.}~\bibnamefont {Lemoult}}, \bibinfo {author}
  {\bibfnamefont {M.}~\bibnamefont {Avila}}, \ and\ \bibinfo {author}
  {\bibfnamefont {B.}~\bibnamefont {Hof}},\ }\href@noop {} {\bibfield
  {journal} {\bibinfo  {journal} {Nature}\ }\textbf {\bibinfo {volume} {526}},\
  \bibinfo {pages} {550} (\bibinfo {year} {2015})}\BibitemShut {NoStop}%
\bibitem [{\citenamefont {Sipos}\ and\ \citenamefont
  {Goldenfeld}(2011)}]{sipos}%
  \BibitemOpen
  \bibfield  {author} {\bibinfo {author} {\bibfnamefont {M.}~\bibnamefont
  {Sipos}}\ and\ \bibinfo {author} {\bibfnamefont {N.}~\bibnamefont
  {Goldenfeld}},\ }\href {\doibase 10.1103/PhysRevE.84.035304} {\bibfield
  {journal} {\bibinfo  {journal} {Phys. Rev. E}\ }\textbf {\bibinfo {volume}
  {84}},\ \bibinfo {pages} {035304} (\bibinfo {year} {2011})}\BibitemShut
  {NoStop}%
\bibitem [{\citenamefont {Barenblatt}(1996)}]{barenblattbook}%
  \BibitemOpen
  \bibfield  {author} {\bibinfo {author} {\bibfnamefont {G.~I.}\ \bibnamefont
  {Barenblatt}},\ }\href@noop {} {\emph {\bibinfo {title} {Scaling,
  self-similarity, and intermediate asymptotics: dimensional analysis and
  intermediate asymptotics}}},\ Vol.~\bibinfo {volume} {14}\ (\bibinfo
  {publisher} {Cambridge University Press},\ \bibinfo {year}
  {1996})\BibitemShut {NoStop}%
\bibitem [{\citenamefont {Davidson}(2000)}]{davidson}%
  \BibitemOpen
  \bibfield  {author} {\bibinfo {author} {\bibfnamefont {P.}~\bibnamefont
  {Davidson}},\ }\href@noop {} {\bibfield  {journal} {\bibinfo  {journal}
  {Journal of Turbulence}\ }\textbf {\bibinfo {volume} {1}},\ \bibinfo {pages}
  {006} (\bibinfo {year} {2000})}\BibitemShut {NoStop}%
\bibitem [{\citenamefont {Hahm}\ and\ \citenamefont {Diamond}(2018)}]{TS18}%
  \BibitemOpen
  \bibfield  {author} {\bibinfo {author} {\bibfnamefont {T.}~\bibnamefont
  {Hahm}}\ and\ \bibinfo {author} {\bibfnamefont {P.}~\bibnamefont {Diamond}},\
  }\href@noop {} {\bibfield  {journal} {\bibinfo  {journal} {Journal of the
  Korean Physical Society}\ }\textbf {\bibinfo {volume} {73}},\ \bibinfo
  {pages} {747} (\bibinfo {year} {2018})}\BibitemShut {NoStop}%
\bibitem [{\citenamefont {Garbet}\ \emph {et~al.}(1994)\citenamefont {Garbet},
  \citenamefont {Laurent}, \citenamefont {Samain},\ and\ \citenamefont
  {Chinardet}}]{garbet94}%
  \BibitemOpen
  \bibfield  {author} {\bibinfo {author} {\bibfnamefont {X.}~\bibnamefont
  {Garbet}}, \bibinfo {author} {\bibfnamefont {L.}~\bibnamefont {Laurent}},
  \bibinfo {author} {\bibfnamefont {A.}~\bibnamefont {Samain}}, \ and\ \bibinfo
  {author} {\bibfnamefont {J.}~\bibnamefont {Chinardet}},\ }\href
  {http://stacks.iop.org/0029-5515/34/i=7/a=I04} {\bibfield  {journal}
  {\bibinfo  {journal} {Nuclear Fusion}\ }\textbf {\bibinfo {volume} {34}},\
  \bibinfo {pages} {963} (\bibinfo {year} {1994})}\BibitemShut {NoStop}%
\bibitem [{\citenamefont {Hahm}\ \emph {et~al.}(2004)\citenamefont {Hahm},
  \citenamefont {Diamond}, \citenamefont {Lin}, \citenamefont {Itoh},\ and\
  \citenamefont {Itoh}}]{hahm04}%
  \BibitemOpen
  \bibfield  {author} {\bibinfo {author} {\bibfnamefont {T.~S.}\ \bibnamefont
  {Hahm}}, \bibinfo {author} {\bibfnamefont {P.~H.}\ \bibnamefont {Diamond}},
  \bibinfo {author} {\bibfnamefont {Z.}~\bibnamefont {Lin}}, \bibinfo {author}
  {\bibfnamefont {K.}~\bibnamefont {Itoh}}, \ and\ \bibinfo {author}
  {\bibfnamefont {S.-I.}\ \bibnamefont {Itoh}},\ }\href
  {http://stacks.iop.org/0741-3335/46/i=5A/a=036} {\bibfield  {journal}
  {\bibinfo  {journal} {Plasma Physics and Controlled Fusion}\ }\textbf
  {\bibinfo {volume} {46}},\ \bibinfo {pages} {A323} (\bibinfo {year}
  {2004})}\BibitemShut {NoStop}%
\bibitem [{\citenamefont {Lin}\ and\ \citenamefont {Hahm}(2004)}]{linhahm}%
  \BibitemOpen
  \bibfield  {author} {\bibinfo {author} {\bibfnamefont {Z.}~\bibnamefont
  {Lin}}\ and\ \bibinfo {author} {\bibfnamefont {T.~S.}\ \bibnamefont {Hahm}},\
  }\href {\doibase 10.1063/1.1647136} {\bibfield  {journal} {\bibinfo
  {journal} {Physics of Plasmas}\ }\textbf {\bibinfo {volume} {11}},\ \bibinfo
  {pages} {1099} (\bibinfo {year} {2004})},\ \Eprint
  {http://arxiv.org/abs/https://doi.org/10.1063/1.1647136}
  {https://doi.org/10.1063/1.1647136} \BibitemShut {NoStop}%
\bibitem [{\citenamefont {Naulin}, \citenamefont {Nielsen},\ and\ \citenamefont
  {Rasmussen}(2005)}]{naulin}%
  \BibitemOpen
  \bibfield  {author} {\bibinfo {author} {\bibfnamefont {V.}~\bibnamefont
  {Naulin}}, \bibinfo {author} {\bibfnamefont {A.~H.}\ \bibnamefont {Nielsen}},
  \ and\ \bibinfo {author} {\bibfnamefont {J.~J.}\ \bibnamefont {Rasmussen}},\
  }\href {\doibase 10.1063/1.2141396} {\bibfield  {journal} {\bibinfo
  {journal} {Physics of Plasmas}\ }\textbf {\bibinfo {volume} {12}},\ \bibinfo
  {pages} {122306} (\bibinfo {year} {2005})},\ \Eprint
  {http://arxiv.org/abs/https://doi.org/10.1063/1.2141396}
  {https://doi.org/10.1063/1.2141396} \BibitemShut {NoStop}%
\bibitem [{\citenamefont {Diamond}\ and\ \citenamefont {Hahm}(1995)}]{DH95}%
  \BibitemOpen
  \bibfield  {author} {\bibinfo {author} {\bibfnamefont {P.~H.}\ \bibnamefont
  {Diamond}}\ and\ \bibinfo {author} {\bibfnamefont {T.~S.}\ \bibnamefont
  {Hahm}},\ }\href {\doibase 10.1063/1.871063} {\bibfield  {journal} {\bibinfo
  {journal} {Physics of Plasmas}\ }\textbf {\bibinfo {volume} {2}},\ \bibinfo
  {pages} {3640} (\bibinfo {year} {1995})},\ \Eprint
  {http://arxiv.org/abs/https://doi.org/10.1063/1.871063}
  {https://doi.org/10.1063/1.871063} \BibitemShut {NoStop}%
\bibitem [{\citenamefont {Politzer}(2000)}]{politzer}%
  \BibitemOpen
  \bibfield  {author} {\bibinfo {author} {\bibfnamefont {P.~A.}\ \bibnamefont
  {Politzer}},\ }\href {\doibase 10.1103/PhysRevLett.84.1192} {\bibfield
  {journal} {\bibinfo  {journal} {Phys. Rev. Lett.}\ }\textbf {\bibinfo
  {volume} {84}},\ \bibinfo {pages} {1192} (\bibinfo {year}
  {2000})}\BibitemShut {NoStop}%
\bibitem [{\citenamefont {Sarazin}\ and\ \citenamefont
  {Ghendrih}(1998)}]{sarazin}%
  \BibitemOpen
  \bibfield  {author} {\bibinfo {author} {\bibfnamefont {Y.}~\bibnamefont
  {Sarazin}}\ and\ \bibinfo {author} {\bibfnamefont {P.}~\bibnamefont
  {Ghendrih}},\ }\href {\doibase 10.1063/1.873157} {\bibfield  {journal}
  {\bibinfo  {journal} {Physics of Plasmas}\ }\textbf {\bibinfo {volume} {5}},\
  \bibinfo {pages} {4214} (\bibinfo {year} {1998})},\ \Eprint
  {http://arxiv.org/abs/https://doi.org/10.1063/1.873157}
  {https://doi.org/10.1063/1.873157} \BibitemShut {NoStop}%
\bibitem [{\citenamefont {Ida}\ \emph {et~al.}(2015)\citenamefont {Ida},
  \citenamefont {Shi}, \citenamefont {Sun}, \citenamefont {Inagaki},
  \citenamefont {Kamiya}, \citenamefont {Rice}, \citenamefont {Tamura},
  \citenamefont {Diamond}, \citenamefont {Dif-Pradalier}, \citenamefont {Zou}
  \emph {et~al.}}]{ida}%
  \BibitemOpen
  \bibfield  {author} {\bibinfo {author} {\bibfnamefont {K.}~\bibnamefont
  {Ida}}, \bibinfo {author} {\bibfnamefont {Z.}~\bibnamefont {Shi}}, \bibinfo
  {author} {\bibfnamefont {H.}~\bibnamefont {Sun}}, \bibinfo {author}
  {\bibfnamefont {S.}~\bibnamefont {Inagaki}}, \bibinfo {author} {\bibfnamefont
  {K.}~\bibnamefont {Kamiya}}, \bibinfo {author} {\bibfnamefont
  {J.}~\bibnamefont {Rice}}, \bibinfo {author} {\bibfnamefont {N.}~\bibnamefont
  {Tamura}}, \bibinfo {author} {\bibfnamefont {P.}~\bibnamefont {Diamond}},
  \bibinfo {author} {\bibfnamefont {G.}~\bibnamefont {Dif-Pradalier}}, \bibinfo
  {author} {\bibfnamefont {X.}~\bibnamefont {Zou}},  \emph {et~al.},\ }\href
  {http://stacks.iop.org/0029-5515/55/i=1/a=013022} {\bibfield  {journal}
  {\bibinfo  {journal} {Nuclear Fusion}\ }\textbf {\bibinfo {volume} {55}},\
  \bibinfo {pages} {013022} (\bibinfo {year} {2015})}\BibitemShut {NoStop}%
\bibitem [{\citenamefont {Inagaki}\ \emph {et~al.}(2013)\citenamefont
  {Inagaki}, \citenamefont {Tokuzawa}, \citenamefont {Tamura}, \citenamefont
  {Itoh}, \citenamefont {Kobayashi}, \citenamefont {Ida}, \citenamefont
  {Shimozuma}, \citenamefont {Kubo}, \citenamefont {Tanaka}, \citenamefont
  {Ido} \emph {et~al.}}]{inagaki}%
  \BibitemOpen
  \bibfield  {author} {\bibinfo {author} {\bibfnamefont {S.}~\bibnamefont
  {Inagaki}}, \bibinfo {author} {\bibfnamefont {T.}~\bibnamefont {Tokuzawa}},
  \bibinfo {author} {\bibfnamefont {N.}~\bibnamefont {Tamura}}, \bibinfo
  {author} {\bibfnamefont {S.-I.}\ \bibnamefont {Itoh}}, \bibinfo {author}
  {\bibfnamefont {T.}~\bibnamefont {Kobayashi}}, \bibinfo {author}
  {\bibfnamefont {K.}~\bibnamefont {Ida}}, \bibinfo {author} {\bibfnamefont
  {T.}~\bibnamefont {Shimozuma}}, \bibinfo {author} {\bibfnamefont
  {S.}~\bibnamefont {Kubo}}, \bibinfo {author} {\bibfnamefont {K.}~\bibnamefont
  {Tanaka}}, \bibinfo {author} {\bibfnamefont {T.}~\bibnamefont {Ido}},  \emph
  {et~al.},\ }\href@noop {} {\bibfield  {journal} {\bibinfo  {journal} {Nuclear
  Fusion}\ }\textbf {\bibinfo {volume} {53}},\ \bibinfo {pages} {113006}
  (\bibinfo {year} {2013})}\BibitemShut {NoStop}%
\bibitem [{\citenamefont {G\"urcan}\ \emph {et~al.}(2005)\citenamefont
  {G\"urcan}, \citenamefont {Diamond}, \citenamefont {Hahm},\ and\
  \citenamefont {Lin}}]{gurcan05}%
  \BibitemOpen
  \bibfield  {author} {\bibinfo {author} {\bibfnamefont {O.~D.}\ \bibnamefont
  {G\"urcan}}, \bibinfo {author} {\bibfnamefont {P.~H.}\ \bibnamefont
  {Diamond}}, \bibinfo {author} {\bibfnamefont {T.~S.}\ \bibnamefont {Hahm}}, \
  and\ \bibinfo {author} {\bibfnamefont {Z.}~\bibnamefont {Lin}},\ }\href
  {\doibase 10.1063/1.1853385} {\bibfield  {journal} {\bibinfo  {journal}
  {Physics of Plasmas}\ }\textbf {\bibinfo {volume} {12}},\ \bibinfo {pages}
  {032303} (\bibinfo {year} {2005})}\BibitemShut {NoStop}%
\bibitem [{\citenamefont {Kim}\ \emph {et~al.}(2003)\citenamefont {Kim},
  \citenamefont {Diamond}, \citenamefont {Malkov}, \citenamefont {Hahm},
  \citenamefont {Itoh}, \citenamefont {Itoh}, \citenamefont {Champeaux},
  \citenamefont {Gruzinov}, \citenamefont {Gurcan},\ and\ \citenamefont
  {Holland}}]{kim}%
  \BibitemOpen
  \bibfield  {author} {\bibinfo {author} {\bibfnamefont {E.}~\bibnamefont
  {Kim}}, \bibinfo {author} {\bibfnamefont {P.}~\bibnamefont {Diamond}},
  \bibinfo {author} {\bibfnamefont {M.}~\bibnamefont {Malkov}}, \bibinfo
  {author} {\bibfnamefont {T.}~\bibnamefont {Hahm}}, \bibinfo {author}
  {\bibfnamefont {K.}~\bibnamefont {Itoh}}, \bibinfo {author} {\bibfnamefont
  {S.-I.}\ \bibnamefont {Itoh}}, \bibinfo {author} {\bibfnamefont
  {S.}~\bibnamefont {Champeaux}}, \bibinfo {author} {\bibfnamefont
  {I.}~\bibnamefont {Gruzinov}}, \bibinfo {author} {\bibfnamefont
  {O.}~\bibnamefont {Gurcan}}, \ and\ \bibinfo {author} {\bibfnamefont
  {C.}~\bibnamefont {Holland}},\ }\href@noop {} {\bibfield  {journal} {\bibinfo
   {journal} {Nuclear Fusion}\ }\textbf {\bibinfo {volume} {43}},\ \bibinfo
  {pages} {961} (\bibinfo {year} {2003})}\BibitemShut {NoStop}%
\bibitem [{\citenamefont {Nazikian}\ \emph {et~al.}(2005)\citenamefont
  {Nazikian}, \citenamefont {Shinohara}, \citenamefont {Kramer}, \citenamefont
  {Valeo}, \citenamefont {Hill}, \citenamefont {Hahm}, \citenamefont {Rewoldt},
  \citenamefont {Ide}, \citenamefont {Koide}, \citenamefont {Oyama} \emph
  {et~al.}}]{nazikian}%
  \BibitemOpen
  \bibfield  {author} {\bibinfo {author} {\bibfnamefont {R.}~\bibnamefont
  {Nazikian}}, \bibinfo {author} {\bibfnamefont {K.}~\bibnamefont {Shinohara}},
  \bibinfo {author} {\bibfnamefont {G.}~\bibnamefont {Kramer}}, \bibinfo
  {author} {\bibfnamefont {E.}~\bibnamefont {Valeo}}, \bibinfo {author}
  {\bibfnamefont {K.}~\bibnamefont {Hill}}, \bibinfo {author} {\bibfnamefont
  {T.}~\bibnamefont {Hahm}}, \bibinfo {author} {\bibfnamefont {G.}~\bibnamefont
  {Rewoldt}}, \bibinfo {author} {\bibfnamefont {S.}~\bibnamefont {Ide}},
  \bibinfo {author} {\bibfnamefont {Y.}~\bibnamefont {Koide}}, \bibinfo
  {author} {\bibfnamefont {Y.}~\bibnamefont {Oyama}},  \emph {et~al.},\
  }\href@noop {} {\bibfield  {journal} {\bibinfo  {journal} {Physical review
  letters}\ }\textbf {\bibinfo {volume} {94}},\ \bibinfo {pages} {135002}
  (\bibinfo {year} {2005})}\BibitemShut {NoStop}%
\bibitem [{\citenamefont {Orszag}\ and\ \citenamefont {Patera}(1983)}]{orszag}%
  \BibitemOpen
  \bibfield  {author} {\bibinfo {author} {\bibfnamefont {S.~A.}\ \bibnamefont
  {Orszag}}\ and\ \bibinfo {author} {\bibfnamefont {A.~T.}\ \bibnamefont
  {Patera}},\ }\href {\doibase 10.1017/S0022112083000518} {\bibfield  {journal}
  {\bibinfo  {journal} {Journal of Fluid Mechanics}\ }\textbf {\bibinfo
  {volume} {128}},\ \bibinfo {pages} {347} (\bibinfo {year}
  {1983})}\BibitemShut {NoStop}%
\bibitem [{\citenamefont {Biskamp}\ and\ \citenamefont
  {Walter}(1985)}]{biskamp}%
  \BibitemOpen
  \bibfield  {author} {\bibinfo {author} {\bibfnamefont {D.}~\bibnamefont
  {Biskamp}}\ and\ \bibinfo {author} {\bibfnamefont {M.}~\bibnamefont
  {Walter}},\ }\href@noop {} {\bibfield  {journal} {\bibinfo  {journal} {Phys.
  Lett.}\ }\textbf {\bibinfo {volume} {109A}} (\bibinfo {year}
  {1985})}\BibitemShut {NoStop}%
\bibitem [{\citenamefont {Drake}, \citenamefont {Zeiler},\ and\ \citenamefont
  {Biskamp}(1995)}]{drake}%
  \BibitemOpen
  \bibfield  {author} {\bibinfo {author} {\bibfnamefont {J.~F.}\ \bibnamefont
  {Drake}}, \bibinfo {author} {\bibfnamefont {A.}~\bibnamefont {Zeiler}}, \
  and\ \bibinfo {author} {\bibfnamefont {D.}~\bibnamefont {Biskamp}},\ }\href
  {\doibase 10.1103/PhysRevLett.75.4222} {\bibfield  {journal} {\bibinfo
  {journal} {Phys. Rev. Lett.}\ }\textbf {\bibinfo {volume} {75}},\ \bibinfo
  {pages} {4222} (\bibinfo {year} {1995})}\BibitemShut {NoStop}%
\bibitem [{\citenamefont {Friedman}\ and\ \citenamefont
  {Carter}(2015)}]{friedman15}%
  \BibitemOpen
  \bibfield  {author} {\bibinfo {author} {\bibfnamefont {B.}~\bibnamefont
  {Friedman}}\ and\ \bibinfo {author} {\bibfnamefont {T.~A.}\ \bibnamefont
  {Carter}},\ }\href {\doibase 10.1063/1.4905863} {\bibfield  {journal}
  {\bibinfo  {journal} {Physics of Plasmas}\ }\textbf {\bibinfo {volume}
  {22}},\ \bibinfo {pages} {012307} (\bibinfo {year} {2015})},\ \Eprint
  {http://arxiv.org/abs/https://doi.org/10.1063/1.4905863}
  {https://doi.org/10.1063/1.4905863} \BibitemShut {NoStop}%
\bibitem [{\citenamefont {Scott}(1990)}]{scott90}%
  \BibitemOpen
  \bibfield  {author} {\bibinfo {author} {\bibfnamefont {B.~D.}\ \bibnamefont
  {Scott}},\ }\href {\doibase 10.1103/PhysRevLett.65.3289} {\bibfield
  {journal} {\bibinfo  {journal} {Phys. Rev. Lett.}\ }\textbf {\bibinfo
  {volume} {65}},\ \bibinfo {pages} {3289} (\bibinfo {year}
  {1990})}\BibitemShut {NoStop}%
\bibitem [{\citenamefont {Scott}(1992)}]{scott92}%
  \BibitemOpen
  \bibfield  {author} {\bibinfo {author} {\bibfnamefont {B.~D.}\ \bibnamefont
  {Scott}},\ }\href {\doibase 10.1063/1.860215} {\bibfield  {journal} {\bibinfo
   {journal} {Physics of Fluids B: Plasma Physics}\ }\textbf {\bibinfo {volume}
  {4}},\ \bibinfo {pages} {2468} (\bibinfo {year} {1992})},\ \Eprint
  {http://arxiv.org/abs/https://doi.org/10.1063/1.860215}
  {https://doi.org/10.1063/1.860215} \BibitemShut {NoStop}%
\bibitem [{\citenamefont {Guo}\ and\ \citenamefont {Diamond}(2017)}]{GD17}%
  \BibitemOpen
  \bibfield  {author} {\bibinfo {author} {\bibfnamefont {Z.~B.}\ \bibnamefont
  {Guo}}\ and\ \bibinfo {author} {\bibfnamefont {P.~H.}\ \bibnamefont
  {Diamond}},\ }\href {\doibase 10.1063/1.5000850} {\bibfield  {journal}
  {\bibinfo  {journal} {Physics of Plasmas}\ }\textbf {\bibinfo {volume} {24}}
  (\bibinfo {year} {2017}),\ 10.1063/1.5000850}\BibitemShut {NoStop}%
\bibitem [{\citenamefont {Dif-Pradalier}\ \emph {et~al.}(2010)\citenamefont
  {Dif-Pradalier}, \citenamefont {Diamond}, \citenamefont {Grandgirard},
  \citenamefont {Sarazin}, \citenamefont {Abiteboul}, \citenamefont {Garbet},
  \citenamefont {Ghendrih}, \citenamefont {Strugarek}, \citenamefont {Ku},\
  and\ \citenamefont {Chang}}]{dif2010}%
  \BibitemOpen
  \bibfield  {author} {\bibinfo {author} {\bibfnamefont {G.}~\bibnamefont
  {Dif-Pradalier}}, \bibinfo {author} {\bibfnamefont {P.~H.}\ \bibnamefont
  {Diamond}}, \bibinfo {author} {\bibfnamefont {V.}~\bibnamefont
  {Grandgirard}}, \bibinfo {author} {\bibfnamefont {Y.}~\bibnamefont
  {Sarazin}}, \bibinfo {author} {\bibfnamefont {J.}~\bibnamefont {Abiteboul}},
  \bibinfo {author} {\bibfnamefont {X.}~\bibnamefont {Garbet}}, \bibinfo
  {author} {\bibfnamefont {P.}~\bibnamefont {Ghendrih}}, \bibinfo {author}
  {\bibfnamefont {A.}~\bibnamefont {Strugarek}}, \bibinfo {author}
  {\bibfnamefont {S.}~\bibnamefont {Ku}}, \ and\ \bibinfo {author}
  {\bibfnamefont {C.~S.}\ \bibnamefont {Chang}},\ }\href {\doibase
  10.1103/PhysRevE.82.025401} {\bibfield  {journal} {\bibinfo  {journal} {Phys.
  Rev. E}\ }\textbf {\bibinfo {volume} {82}},\ \bibinfo {pages} {025401}
  (\bibinfo {year} {2010})}\BibitemShut {NoStop}%
\bibitem [{\citenamefont {Carreras}\ \emph {et~al.}(1992)\citenamefont
  {Carreras}, \citenamefont {Sidikman}, \citenamefont {Diamond}, \citenamefont
  {Terry},\ and\ \citenamefont {Garcia}}]{CSD92}%
  \BibitemOpen
  \bibfield  {author} {\bibinfo {author} {\bibfnamefont {B.~A.}\ \bibnamefont
  {Carreras}}, \bibinfo {author} {\bibfnamefont {K.}~\bibnamefont {Sidikman}},
  \bibinfo {author} {\bibfnamefont {P.~H.}\ \bibnamefont {Diamond}}, \bibinfo
  {author} {\bibfnamefont {P.~W.}\ \bibnamefont {Terry}}, \ and\ \bibinfo
  {author} {\bibfnamefont {L.}~\bibnamefont {Garcia}},\ }\href {\doibase
  10.1063/1.860420} {\bibfield  {journal} {\bibinfo  {journal} {Physics of
  Fluids B: Plasma Physics}\ }\textbf {\bibinfo {volume} {4}},\ \bibinfo
  {pages} {3115} (\bibinfo {year} {1992})},\ \Eprint
  {http://arxiv.org/abs/https://doi.org/10.1063/1.860420}
  {https://doi.org/10.1063/1.860420} \BibitemShut {NoStop}%
\bibitem [{\citenamefont {Pringle}, \citenamefont {McMillan},\ and\
  \citenamefont {Teaca}(2017)}]{pringle}%
  \BibitemOpen
  \bibfield  {author} {\bibinfo {author} {\bibfnamefont {C.~C.~T.}\
  \bibnamefont {Pringle}}, \bibinfo {author} {\bibfnamefont {B.~F.}\
  \bibnamefont {McMillan}}, \ and\ \bibinfo {author} {\bibfnamefont
  {B.}~\bibnamefont {Teaca}},\ }\href {\doibase 10.1063/1.4999848} {\bibfield
  {journal} {\bibinfo  {journal} {Physics of Plasmas}\ }\textbf {\bibinfo
  {volume} {24}},\ \bibinfo {pages} {122307} (\bibinfo {year} {2017})},\
  \Eprint {http://arxiv.org/abs/https://doi.org/10.1063/1.4999848}
  {https://doi.org/10.1063/1.4999848} \BibitemShut {NoStop}%
\bibitem [{\citenamefont {Barnes}\ \emph {et~al.}(2011)\citenamefont {Barnes},
  \citenamefont {Parra}, \citenamefont {Highcock}, \citenamefont
  {Schekochihin}, \citenamefont {Cowley},\ and\ \citenamefont
  {Roach}}]{barnes}%
  \BibitemOpen
  \bibfield  {author} {\bibinfo {author} {\bibfnamefont {M.}~\bibnamefont
  {Barnes}}, \bibinfo {author} {\bibfnamefont {F.~I.}\ \bibnamefont {Parra}},
  \bibinfo {author} {\bibfnamefont {E.~G.}\ \bibnamefont {Highcock}}, \bibinfo
  {author} {\bibfnamefont {A.~A.}\ \bibnamefont {Schekochihin}}, \bibinfo
  {author} {\bibfnamefont {S.~C.}\ \bibnamefont {Cowley}}, \ and\ \bibinfo
  {author} {\bibfnamefont {C.~M.}\ \bibnamefont {Roach}},\ }\href {\doibase
  10.1103/PhysRevLett.106.175004} {\bibfield  {journal} {\bibinfo  {journal}
  {Phys. Rev. Lett.}\ }\textbf {\bibinfo {volume} {106}},\ \bibinfo {pages}
  {175004} (\bibinfo {year} {2011})}\BibitemShut {NoStop}%
\bibitem [{\citenamefont {van Wyk}\ \emph {et~al.}(2016)\citenamefont {van
  Wyk}, \citenamefont {Highcock}, \citenamefont {Schekochihin}, \citenamefont
  {Roach}, \citenamefont {Field},\ and\ \citenamefont {Dorland}}]{vanwyk}%
  \BibitemOpen
  \bibfield  {author} {\bibinfo {author} {\bibfnamefont {F.}~\bibnamefont {van
  Wyk}}, \bibinfo {author} {\bibfnamefont {E.~G.}\ \bibnamefont {Highcock}},
  \bibinfo {author} {\bibfnamefont {A.~A.}\ \bibnamefont {Schekochihin}},
  \bibinfo {author} {\bibfnamefont {C.~M.}\ \bibnamefont {Roach}}, \bibinfo
  {author} {\bibfnamefont {A.~R.}\ \bibnamefont {Field}}, \ and\ \bibinfo
  {author} {\bibfnamefont {W.}~\bibnamefont {Dorland}},\ }\href {\doibase
  10.1017/S0022377816001148} {\bibfield  {journal} {\bibinfo  {journal}
  {Journal of Plasma Physics}\ }\textbf {\bibinfo {volume} {82}},\ \bibinfo
  {pages} {905820609} (\bibinfo {year} {2016})}\BibitemShut {NoStop}%
\bibitem [{\citenamefont {Terry}, \citenamefont {Diamond},\ and\ \citenamefont
  {Hahm}(1990)}]{holes}%
  \BibitemOpen
  \bibfield  {author} {\bibinfo {author} {\bibfnamefont {P.~W.}\ \bibnamefont
  {Terry}}, \bibinfo {author} {\bibfnamefont {P.~H.}\ \bibnamefont {Diamond}},
  \ and\ \bibinfo {author} {\bibfnamefont {T.~S.}\ \bibnamefont {Hahm}},\
  }\href {\doibase 10.1063/1.859426} {\bibfield  {journal} {\bibinfo  {journal}
  {Physics of Fluids B: Plasma Physics}\ }\textbf {\bibinfo {volume} {2}},\
  \bibinfo {pages} {2048} (\bibinfo {year} {1990})},\ \Eprint
  {http://arxiv.org/abs/https://doi.org/10.1063/1.859426}
  {https://doi.org/10.1063/1.859426} \BibitemShut {NoStop}%
\bibitem [{\citenamefont {Lesur}\ and\ \citenamefont {Diamond}(2013)}]{lesur}%
  \BibitemOpen
  \bibfield  {author} {\bibinfo {author} {\bibfnamefont {M.}~\bibnamefont
  {Lesur}}\ and\ \bibinfo {author} {\bibfnamefont {P.~H.}\ \bibnamefont
  {Diamond}},\ }\href {\doibase 10.1103/PhysRevE.87.031101} {\bibfield
  {journal} {\bibinfo  {journal} {Phys. Rev. E}\ }\textbf {\bibinfo {volume}
  {87}},\ \bibinfo {pages} {031101} (\bibinfo {year} {2013})}\BibitemShut
  {NoStop}%
\bibitem [{\citenamefont {Gil}\ and\ \citenamefont
  {Sornette}(1996)}]{gilsornette}%
  \BibitemOpen
  \bibfield  {author} {\bibinfo {author} {\bibfnamefont {L.}~\bibnamefont
  {Gil}}\ and\ \bibinfo {author} {\bibfnamefont {D.}~\bibnamefont {Sornette}},\
  }\href {\doibase 10.1103/PhysRevLett.76.3991} {\bibfield  {journal} {\bibinfo
   {journal} {Phys. Rev. Lett.}\ }\textbf {\bibinfo {volume} {76}},\ \bibinfo
  {pages} {3991} (\bibinfo {year} {1996})}\BibitemShut {NoStop}%
\bibitem [{\citenamefont {Zel'dovich}\ and\ \citenamefont
  {Frank-Kamenetsky}(1938)}]{ZFK}%
  \BibitemOpen
  \bibfield  {author} {\bibinfo {author} {\bibfnamefont {Y.}~\bibnamefont
  {Zel'dovich}}\ and\ \bibinfo {author} {\bibfnamefont {D.}~\bibnamefont
  {Frank-Kamenetsky}},\ }in\ \href@noop {} {\emph {\bibinfo {booktitle}
  {Doklady AN SSSR}}},\ Vol.~\bibinfo {volume} {19}\ (\bibinfo {year} {1938})\
  pp.\ \bibinfo {pages} {693--697}\BibitemShut {NoStop}%
\bibitem [{\citenamefont {S{\'a}nchez-Gardu{\~{n}}o}\ and\ \citenamefont
  {Maini}(1997)}]{SGM97}%
  \BibitemOpen
  \bibfield  {author} {\bibinfo {author} {\bibfnamefont {F.}~\bibnamefont
  {S{\'a}nchez-Gardu{\~{n}}o}}\ and\ \bibinfo {author} {\bibfnamefont {P.~K.}\
  \bibnamefont {Maini}},\ }\href {\doibase 10.1007/s002850050073} {\bibfield
  {journal} {\bibinfo  {journal} {Journal of Mathematical Biology}\ }\textbf
  {\bibinfo {volume} {35}},\ \bibinfo {pages} {713} (\bibinfo {year}
  {1997})}\BibitemShut {NoStop}%
\bibitem [{\citenamefont {Ghendrih}\ \emph {et~al.}(2007)\citenamefont
  {Ghendrih}, \citenamefont {Sarazin}, \citenamefont {Ciraolo}, \citenamefont
  {Darmet}, \citenamefont {Garbet}, \citenamefont {Grangirard}, \citenamefont
  {Tamain}, \citenamefont {Benkadda},\ and\ \citenamefont {Beyer}}]{ghendrih}%
  \BibitemOpen
  \bibfield  {author} {\bibinfo {author} {\bibfnamefont {P.}~\bibnamefont
  {Ghendrih}}, \bibinfo {author} {\bibfnamefont {Y.}~\bibnamefont {Sarazin}},
  \bibinfo {author} {\bibfnamefont {G.}~\bibnamefont {Ciraolo}}, \bibinfo
  {author} {\bibfnamefont {G.}~\bibnamefont {Darmet}}, \bibinfo {author}
  {\bibfnamefont {X.}~\bibnamefont {Garbet}}, \bibinfo {author} {\bibfnamefont
  {V.}~\bibnamefont {Grangirard}}, \bibinfo {author} {\bibfnamefont
  {P.}~\bibnamefont {Tamain}}, \bibinfo {author} {\bibfnamefont
  {S.}~\bibnamefont {Benkadda}}, \ and\ \bibinfo {author} {\bibfnamefont
  {P.}~\bibnamefont {Beyer}},\ }\href@noop {} {\bibfield  {journal} {\bibinfo
  {journal} {Journal of nuclear materials}\ }\textbf {\bibinfo {volume}
  {363}},\ \bibinfo {pages} {581} (\bibinfo {year} {2007})}\BibitemShut
  {NoStop}%
\bibitem [{\citenamefont {Pomeau}\ and\ \citenamefont
  {Manneville}(1980)}]{pomeau80}%
  \BibitemOpen
  \bibfield  {author} {\bibinfo {author} {\bibfnamefont {Y.}~\bibnamefont
  {Pomeau}}\ and\ \bibinfo {author} {\bibfnamefont {P.}~\bibnamefont
  {Manneville}},\ }\href {https://projecteuclid.org:443/euclid.cmp/1103907981}
  {\bibfield  {journal} {\bibinfo  {journal} {Comm. Math. Phys.}\ }\textbf
  {\bibinfo {volume} {74}},\ \bibinfo {pages} {189} (\bibinfo {year}
  {1980})}\BibitemShut {NoStop}%
\bibitem [{\citenamefont {Van~Compernolle}\ \emph {et~al.}(2015)\citenamefont
  {Van~Compernolle}, \citenamefont {Morales}, \citenamefont {Maggs},\ and\
  \citenamefont {Sydora}}]{compernolle}%
  \BibitemOpen
  \bibfield  {author} {\bibinfo {author} {\bibfnamefont {B.}~\bibnamefont
  {Van~Compernolle}}, \bibinfo {author} {\bibfnamefont {G.~J.}\ \bibnamefont
  {Morales}}, \bibinfo {author} {\bibfnamefont {J.~E.}\ \bibnamefont {Maggs}},
  \ and\ \bibinfo {author} {\bibfnamefont {R.~D.}\ \bibnamefont {Sydora}},\
  }\href {\doibase 10.1103/PhysRevE.91.031102} {\bibfield  {journal} {\bibinfo
  {journal} {Phys. Rev. E}\ }\textbf {\bibinfo {volume} {91}},\ \bibinfo
  {pages} {031102} (\bibinfo {year} {2015})}\BibitemShut {NoStop}%
\end{thebibliography}

\end{document}